\documentclass[pdflatex,sn-mathphys-num]{sn-jnl}

\usepackage{graphicx}%
\usepackage{multirow}%
\usepackage{amsmath,amssymb,amsfonts}%
\usepackage{amsthm}%
\usepackage{mathrsfs}%
\usepackage[title]{appendix}%
\usepackage{xcolor}%
\usepackage{textcomp}%
\usepackage{manyfoot}%
\usepackage{booktabs}%
\usepackage{algorithm}%
\usepackage{algorithmicx}%
\usepackage{algpseudocode}%
\usepackage{listings}%
\usepackage{gensymb} 
\usepackage{hyperref}

\theoremstyle{thmstyleone}%
\theoremstyle{thmstyletwo}%

\theoremstyle{thmstylethree}%

\begin{document}

\title[Article Title]{Current State of Atmospheric Turbulence Cascades}


\author*[1]{\fnm{Vicente} \sur{Corral Arreola}}\email{vinod.kumar@tamuk.edu}
\equalcont{These authors contributed equally to this work.}

\author[1]{\fnm{Arturo} \sur{Rodriguez}}\email{arodriguez123@miners.utep.edu}
\equalcont{These authors contributed equally to this work.}

\author[2]{\fnm{Vinod} \sur{Kumar}}\email{vcorral7@miners.utep.edu}
\equalcont{These authors contributed equally to this work.}

\affil*[1]{\orgdiv{Department of Aerospace and Mechanical Engineering}, \orgname{The University of Texas at El Paso}, \orgaddress{\street{500 W University Ave}, \city{El Paso}, \postcode{79968}, \state{Texas}, \country{United States}}}

\affil[2]{\orgdiv{Department of Mechanical and Industrial Engineering}, \orgname{Texas A\&M University at Kingsville}, \orgaddress{\street{700 University Blvd}, \city{Kingsville}, \postcode{78363}, \state{Texas}, \country{United States}}}


\abstract{Turbulence cascade has been modeled using various methods; the one we have used applies to a more exact representation of turbulence where people use the multifractal representation. The nature of the energy dissipation is usually governed by partial differential equations that have been described, such as Navier-Stokes Equations, although usually in climate modeling, the Kolmogorov turbulence cascading approximation leads towards an isotropic representation. In recent years, Meneveau et al. have proposed to go away from Kolmogorov assumptions and propose multifractal models where we can account for a new anisotropic representation. Our research aims to use Direct Numerical Simulations (DNS) from JHU Turbulence Database and Large Eddy Simulations (LES) we simulated using OpenFOAM to predict how accurate these simulations are in replicating Meneveau experimental procedures with numerical simulations using the same rigorous mathematical approaches. Modeling turbulence cascading using higher fidelity data will advance the field and produce faster and better remote sensing metrics. We have written computer code to analyze DNS and LES data and study the multifractal nature of energy dissipation. The box-counting method is used to identify the multifractal dimension spectrum of the DNS and LES data in every direction to follow Meneveau’s work to represent turbulence-cascading effects in the atmosphere better.}




\maketitle

\section{Introduction}\label{sec1}

Atmospheric turbulence refers to the random fluctuations of the atmosphere’s fluid flow. Turbulence is characterized by chaotic changes in pressure and flow velocity. Many practical problems/applications require an accurate description/model of the atmosphere and how atmospheric turbulence evolves over time\cite{bib1}. Most seen in weather prediction but also applied in remote sensing for defense projects and boundary layers on surfaces\cite{bib2}. Three of the most known simulations currently employed are Direct Numerical Simulation (DNS), Large-Eddy Simulations (LES), and Reynolds-Averaged Navier Stokes (RANS). Each provides a different level of fidelity which offers the ability to produce quick approximations when a higher fidelity simulation would take longer time to approximate. There are various methods that look at different parameters of atmospheric turbulence that allow for a different description of the turbulence \cite{bib3, bib4, bib5, bib6}. However, because of the chaotic nature of these systems, the approximations will lose their accuracy after some time. This is the root problem for any scientist looking at turbulent fluid dynamics and why many explore/implement different methods when analysing such systems. 
 \\ \\
Fluid dynamics are usually described as eddies or vertices occurring in the fluid flow. Usually, an eddy is an area of homogenous velocity, and thus eddies have a length. The dissipation of the eddies refers to the decay of eddies from large eddies to small eddies. Often, one will look at a simpler problem in the search for a pattern that can then be extended to a more complex and practical problem. In fluid dynamics, isotropic descriptions are easier to model because it provides the constraint that eddies, their velocity, and thus their energy dissipates equally along all three dimensions \cite{bib7}. This is in contrast with what we observe in nature, called anisotropic. Kolmogorov came up with a power law that describes isotropic turbulent flow, known as the -5/3 power law, which governs the dissipation in the inertial range. This power law has been expanded using multiplicative processes and multifractals to describe anisotropic dissipation \cite{bib3}.

\section{Literature Review}\label{sec2}


One of Kolmogorov's most important contributions is the development of the 5/3 power scaling law, which provides a fundamental description of the energy distribution in turbulent flows. Kolmogorov's 5/3 power scaling law is derived from three hypotheses he proposed in 1941, which aimed to describe the statistical properties of turbulent flows.
\\ \\
\textbf{Homogeneity and Isotropy:} Kolmogorov assumed that, at sufficiently small scales, the flow is statistically homogeneous and isotropic. This means that the flow properties are invariant under translations and rotations, making them independent of spatial orientation or position.
\\ \\
\textbf{Locality:} Kolmogorov postulated that energy transfer in a turbulent flow is a local process, meaning that energy cascades from larger scales to smaller scales without directly influencing scales much larger or much smaller than itself.
\\ \\
\textbf{Self-similarity:} Kolmogorov hypothesized that the statistics of the small-scale turbulent motion are universal and depend only on the rate of energy dissipation per unit mass ($\varepsilon$) and the kinematic viscosity of the fluid ($\nu$).

\begin{figure}[h!]
    \centering
    \includegraphics[width=0.8\textwidth]{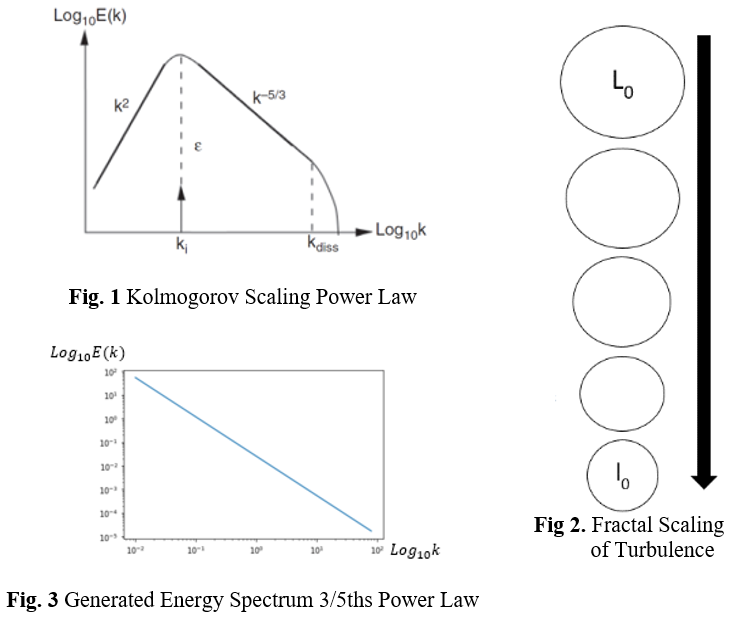} 
\end{figure}

\noindent The power spectrum, denoted as $E(k)$, represents the energy content of the turbulent flow as a function of the wavenumber $k$ (which is inversely proportional to the length scale). In the so-called ``inertial subrange'' of scales, where the energy cascade is predominantly local and unaffected by the viscous dissipation, Kolmogorov found that the power spectrum follows a specific scaling relationship:

\begin{equation}
E(k) \sim \varepsilon^{2/3} k^{-5/3}
\end{equation}

\begin{equation}
k = \frac{2\pi}{\lambda}
\end{equation}

\noindent This relationship, known as Kolmogorov's 5/3 power scaling law, implies that the energy of turbulent fluctuations decreases with increasing wavenumber (or decreasing length scale) following a power law with an exponent of -5/3. The law holds in the inertial subrange, which lies between the energy-containing range (larger scales, where energy is injected into the flow) and the dissipative range (smaller scales, where energy is dissipated by viscosity). Kolmogorov's 5/3 power scaling law has had a profound impact on our understanding of turbulence. It provides a simple yet powerful framework for characterizing the complex, multiscale nature of turbulent flows, and it has been widely validated by experimental and computational studies. Multifractal turbulence unifies understanding of turbulent flows across various scales in space and time. Turbulence comprises self-similar, multifractal structures with diverse statistical scaling behaviors resulting from energy cascade and intermittency processes. The multifractal formalism enables quantitative characterization of turbulence's complex geometry and dynamics, uncovering scaling relationships and fluctuation patterns. This approach helps researchers gain deeper insights into turbulent phenomena and the mechanisms driving anisotropic fluid motion and energy dissipation across diverse scales \cite{bib8, bib9, bib10, bib11}.

\begin{figure}[h!]
    \centering
    \includegraphics[width=1\textwidth]{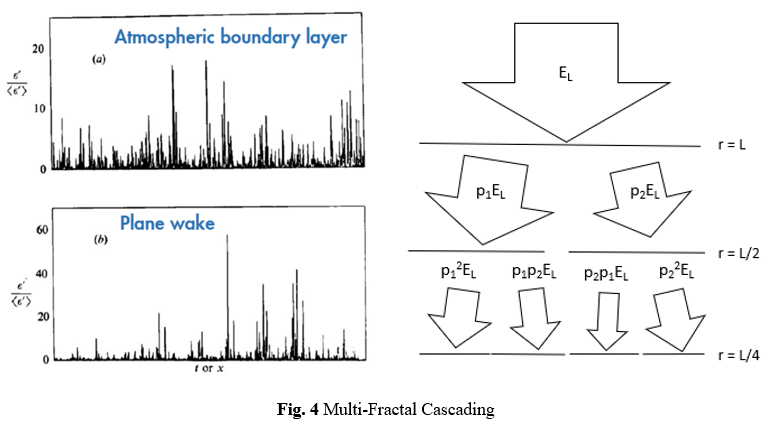} 
    \par\vspace{0.5em}
\end{figure}

\section{Methodology}\label{sec3}

\subsection{Johns Hopkins Turbulence Database (JHTDB)}\label{subsec3}

The Johns Hopkins Turbulence Database (JHTDB) is a comprehensive collection of high-fidelity data sets for studying turbulence in fluid dynamics. Researchers at Johns Hopkins University created the JHTDB, widely used by scientists and engineers worldwide to model and simulate turbulence in various fields. The JHTDB provides several types of data, including velocity, pressure, and vorticity fields, as well as spectral and statistical information. The data sets are generated using Direct Numerical Simulation (DNS), a highly accurate computational method for solving the equations that describe fluid flow. The JHTDB is available to researchers and includes a user-friendly web interface that allows users to search and download data sets. The database also provides tools for visualizing and analyzing the data, including software for computing statistical quantities and generating animations. The JHTDB is an essential resource for researchers in fluid dynamics as it provides highly accurate data for validating models and simulations of turbulence. The database is used to develop new turbulence models and investigate the fundamental properties of turbulence. The JHTDB has been used in various research applications, including studying turbulent mixing in combustion processes, developing turbulence models for large-eddy simulations, and analyzing the turbulent boundary layer in high-speed flows. In summary, the JHTDB is a valuable and widely used resource for researchers in fluid dynamics, providing highly accurate data and tools for analyzing and visualizing turbulence. 

\begin{figure}[h!]
    \centering
    \includegraphics[width=1\textwidth]{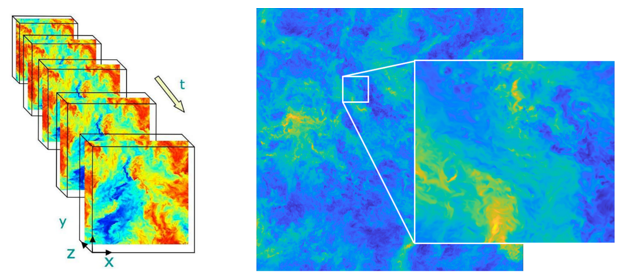} 
    \par\vspace{0.5em}
    \textbf{Fig. 5} Johns Hopkins Turbulence Database Snapshots 
\end{figure}

\subsection{Multi-Fractal Model Implementing Multiplicative Process and Calculations}\label{subsec3}

In 1991, Meneveau introduced a new approach to modelling turbulence based on the concept of multifractals. The multifractal approach is based on the observation that turbulence is not self-similar, meaning that the statistical properties of turbulence change with length scale. Multifractal models describe the scaling properties of turbulence at different length scales and consider that other flow regions may have different scaling behaviours. Meneveau's paper introduced a new multifractal model for turbulence, which was based on the idea of using a probability density function to describe the distribution of turbulent fluctuations at different length scales. The model accurately captured the statistical properties of turbulence in both the inertial and dissipative ranges. As in the Journal of Fluid Mechanics paper published in 1991, Charles Meneveau introduced a new approach to modeling turbulence based on multiplicative processes \cite{bib8}. The equations for the multiplicative processes can be described as follows.
\\ \\
First, the velocity field can be decomposed into a set of scaling coefficients, $C(q)$, and a set of spatial structures, $\varphi(x)$, using the following equation:

\begin{equation}
u(x) = \sum C(q)\varphi(q, x)
\end{equation}
\\ 
where $q$ is the scaling exponent and $x$ is the spatial location. Next, the scaling coefficients, $C(q)$, are assumed to follow a multiplicative process, which can be expressed as:

\begin{equation}
C(q) = \prod \chi(i, q)
\end{equation}
\\ \\
where $\chi(i, q)$ are the local scaling exponents.
\\ \\
The probability density function of the velocity fluctuations can be derived using the multiplicative process and is given by:
\begin{equation}
P(u) = \prod \psi(i)(u(i))
\end{equation}
where $\psi(i)(u(i))$ is the probability density function of the local velocity fluctuations.
\\ \\
Overall, these equations describe Meneveau's multiplicative process model, which is a powerful tool for understanding the statistical properties of turbulence. 

\begin{figure}[h!]
    \centering
    \includegraphics[width=1\textwidth]{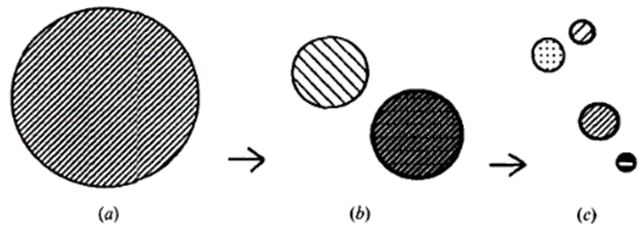} 
    \par\vspace{0.5em}
    \textbf{Fig. 6} Multiplicative Processes Model 
\end{figure}

\subsection{Python Program for Mesh and Energy Dissipation}\label{subsec3}

Accessing the isotropic1024 fine dataset via a web browser allows one to retrieve various profiles at specific points in space and time; specifically, the velocity was retrieved. By uploading a CSV file that contains a discretized mesh, one can get the velocities in bulk; however, the time must be changed manually; a $\Delta t$ of 0.002 seconds was used for the entirety of the 0.198-second run available. The mesh was created using a Python program that created a 32 by 32 by 32 mesh with 0.01 intervals to obtain the velocities across all positive directions.
\begin{equation}
\frac{\varepsilon'}{\langle \varepsilon \rangle}, \quad \varepsilon' = \frac{\partial u_1}{\partial t}, \quad \langle \varepsilon \rangle = \frac{\sum \sum u_{ij}}{t}
\end{equation}
The energy dissipation given by $\varepsilon'$ uses the velocity in one direction. Tracking the dissipation along the positive $x$ direction, it was normalized by the average of all the points, $i$ and $j$, and across all the time steps $\Delta t$.

\subsection{OpenFOAM Large-Eddy Simulations}\label{subsec3}

In this study, we have used OpenFOAM to simulate atmospheric turbulence using the Large Eddy Simulation (LES) model, where we take a cube where the turbulence dissipates isotropically. Knowing that turbulence dissipates isotropically, we wonder if, using the methods described by Meneveau, the fractal dimension would be the same in all directions. We took snapshots at two different times to compare unsteadiness and coherence in turbulence, and we found that all directions have the same structure. We realized that the fractal dimension is the same \cite{bib12}. Large-Eddy Simulations can be treated as a low pass filter which is beneficial in resolving the middle range scales, which works for typical turbulence. Although there are different turbulence levels, a more refined model for deep turbulence will be needed where Direct Numerical Simulations (DNS) can be helpful and necessary. For incompressible flow, which is our case, the LES Navier-Stokes equations are described as follows.
\\ \\
The Continuity Equation:
\begin{equation}
\frac{\partial \overline{u}_l}{\partial x_i} = 0
\end{equation}
The Momentum Equations in Einstein Notation:
\begin{equation}
\frac{\partial \overline{u}_l}{\partial t} + \frac{\partial}{\partial x_j} \left( u_l u_j \right) = 
-\frac{1}{\rho} \frac{\partial \overline{p}}{\partial x_i} + \nu \frac{\partial}{\partial x_j} \left( \frac{\partial \overline{u}_l}{\partial x_j} + \frac{\partial \overline{u}_j}{\partial x_i} \right) = 
-\frac{1}{\rho} \frac{\partial \overline{p}}{\partial x_i} + 2 \nu \frac{\partial}{\partial x_j} \overline{S}_{ij}
\end{equation}

\subsection{Box Counting}\label{subsec3}

Fractals are intricate, self-similar patterns found throughout nature and mathematics. These geometric shapes possess fascinating properties, and their study has led to significant advancements in various fields, including the understanding of turbulent flows. One technique for analyzing fractals, box-counting, has proven especially useful in investigating the complex phenomenon of turbulence. Box-counting is a method used to quantify the scaling behavior of fractals. It is based on the idea of covering a fractal with a grid of small, non-overlapping boxes and then counting the number of boxes required to cover the entire fractal. This process is repeated for different box sizes, and the relationship between the number of boxes and the box size is analyzed to determine the fractal dimension. Fractal dimension (D) measures how a fractal fills space or scales with magnification. For simple geometric shapes like a line or a plane, the dimension is an integer (e.g., 1 for a line, 2 for a plane). However, fractals often have non-integer dimensions, which indicate their complexity and self-similarity across different scales. The box-counting dimension is a specific type of fractal dimension that can be calculated as follows:

\begin{equation}
D = -\lim \frac{\log N}{\log s}
\end{equation}
\\ \\
Where N is the number of boxes required to cover the fractal, s is the side length of the box, and the limit is taken as s approaches zero\cite{bib13}.

\begin{figure}[h!]
    \centering
    \includegraphics[width=1\textwidth]{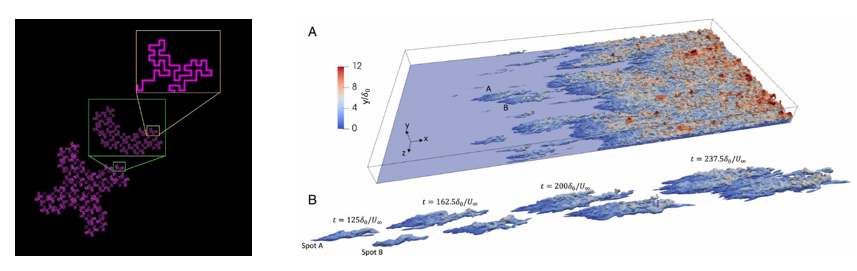} 
    \par\vspace{0.5em}
    \textbf{Fig. 7} Fractal Nature of Turbulence 
\end{figure}

\begin{figure}[h!]
    \centering
    \includegraphics[width=0.70\textwidth]{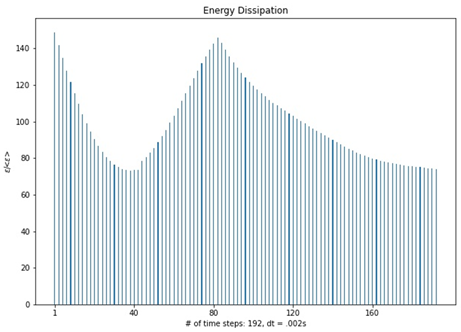} 
    \par\vspace{0.5em}
    \textbf{Fig. 8} Temporal Energy Dissipation 
\end{figure}

\section{Results}\label{sec3}

This bar plot shows the energy dissipation in terms of time. As we have discussed, we have used OpenFOAM to simulate atmospheric turbulence where these images captured the velocity magnitude in time.

\begin{figure}[h!]
    \centering
    \includegraphics[width=0.9\textwidth]{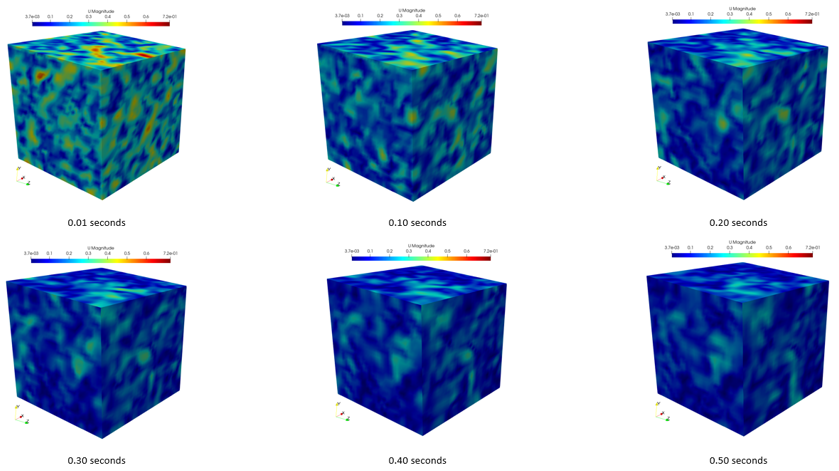} 
    \par\vspace{0.5em}
    \textbf{Fig. 9} Temporal Energy Dissipation using OpenFOAM Large-Eddy Simulations 
\end{figure}
\noindent We see fractal agreement and consistency at 0.01s for isotropic Large-Eddy Simulations.
\begin{figure}[h!]
    \centering
    \includegraphics[width=0.9\textwidth]{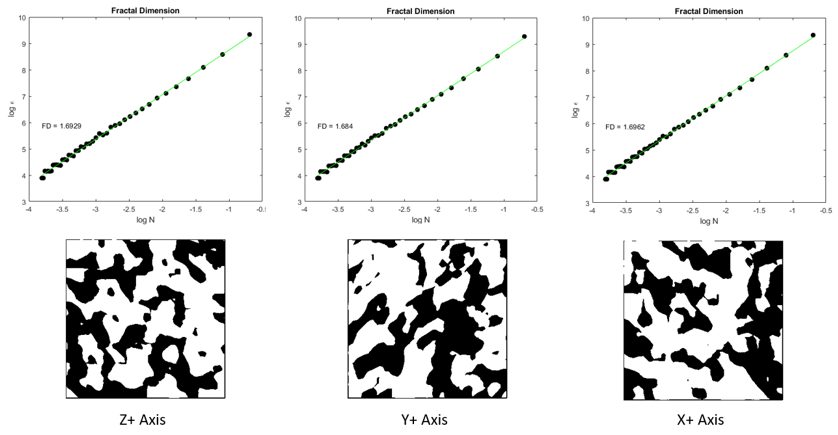} 
    \par\vspace{0.5em}
    \textbf{Fig. 10} Fractal Nature at 0.01s 
\end{figure}
We see fractal agreement and consistency at 0.40s for isotropic Large-Eddy Simulations.
\begin{figure}[h!]
    \centering
    \includegraphics[width=0.9\textwidth]{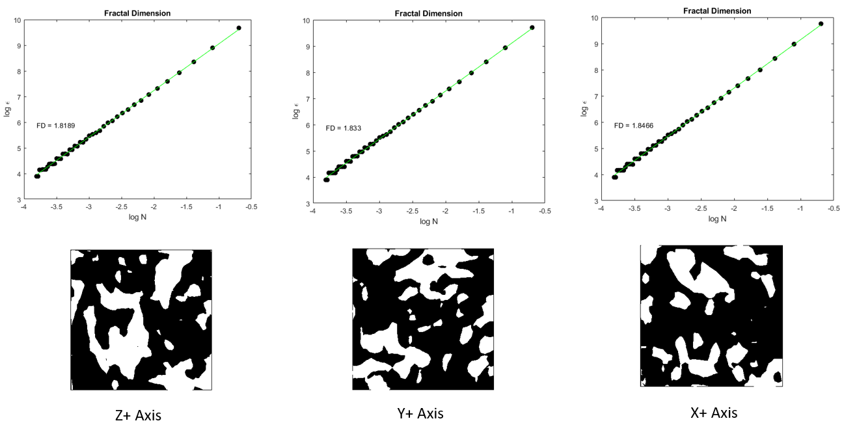} 
    \par\vspace{0.5em}
    \textbf{Fig. 11} Fractal Nature at 0.40s 
\end{figure}

\section{Conclusions}\label{sec5}
We conducted multifractal turbulence analysis on various turbulence datasets and discovered that all simulated datasets exhibit isotropic characteristics. In agreement with Meneveau's findings, our simulations demonstrated that the box-counting fractal dimension remains consistent across all directions. This highlights the need for incorporating experimental data into our analysis for a more comprehensive multifractal understanding. Future research should focus on devising methods to perform multifractal turbulent simulations, drawing inspiration from experimental data, as demonstrated in Meneveau's work, to overcome the limitations imposed by isotropic DNS and LES datasets. 

\section*{Acknowledgments}
We acknowledge Kate Reza and Rene D. Reza for helping with the JHU Database, the U.S. Department of Defense (AFOSR Grant Number \# FA9550-19-1-0304, FA9550-17-1-0253, FA9550-12-1-0242, FA9550-17-1-0393, SFFP, AFTC, HAFB/HSTT, AFRL, HPCMP), U.S. Department of Energy (GRANT13584020, DE-SC0022957, DE-FE0026220, DE-FE0002407, NETL, Sandia, ORNL, NREL), Systems Plus, and several other individuals at these agencies for partially supporting our research financially or through mentorship. We would also like to thank NSF ((HRD-1139929, XSEDE Award Number ACI-1053575), TACC, DOE, DOD, HPCMP, University of Texas STAR program, UTEP (Research Cloud, Department of Mechanical Engineering, Graduate School \& College of Engineering) for generously providing financial support or computational resources. Without their generous support, completing the milestones would have been almost impossible.

\end{document}